\documentclass[final,sort]{aipproc}

\layoutstyle{8x11double}

\newcommand{\aap}{A\&A}
\newcommand{\aj}{AJ}
\newcommand{\apj}{ApJ}
\newcommand{\apjl}{ApJ}
\newcommand{\apjs}{ApJS}
\newcommand{\araa}{ARA\&A}
\newcommand{\mnras}{MNRAS}
\newcommand{\nat}{Nature}
\newcommand{\physrep}{Phys. Rep.}

\begin{document}

\title{Towards the First Galaxies}



\author{Thomas H. Greif}{address={Institut f\"{u}r Theoretische Astrophysik, Albert-Ueberle Strasse 2, 69120 Heidelberg, Germany},altaddress={Department of Astronomy, University of Texas, Austin, TX 78712}}

\author{Jarrett L. Johnson}{address={Department of Astronomy, University of Texas, Austin, TX 78712}}

\author{Volker Bromm}{address={Department of Astronomy, University of Texas, Austin, TX 78712}}

\begin{abstract}
The formation of the first galaxies at redshifts $z\sim 10-15$ signaled the transition from the simple initial state of the universe to one of ever increasing complexity. We here review recent progress in understanding their assembly process with numerical simulations, starting with cosmological initial conditions and modelling the detailed physics of star formation. In particular, we study the role of HD cooling in ionized primordial gas, the impact of UV radiation produced by the first stars, and the propagation of the supernova blast waves triggered at the end of their brief lives. We conclude by discussing promising observational diagnostics that will allow us to probe the properties of the first galaxies, such as their contribution to reionization and the chemical abundance pattern observed in extremely low-metallicity stars.
\end{abstract}

\maketitle

\section{Introduction}
One of the key goals in modern cosmology is to study the assembly process of the first galaxies, and understand how the first stars, stellar systems, and massive black holes formed at the end of the cosmic dark ages, a few hundred million years after the Big Bang. With the formation of the first stars, the so-called Population~III (Pop~III), the universe was rapidly transformed into an increasingly complex, hierarchical system, due to the energy and heavy element input from the first stars and accreting black holes \cite{bl01,bl04a,cf05,miralda03}. Currently, we can directly probe the state of the universe roughly a million years after the Big Bang by detecting the anisotropies in the cosmic microwave background (CMB), thus providing us with the initial conditions for subsequent structure formation. Complementary to the CMB observations, we can probe cosmic history all the way from the present-day universe to roughly a billion years after the Big Bang, using the best available ground- and space-based telescopes. In between lies the remaining frontier, and the first galaxies are the sign-posts of this early, formative epoch.

The tight correlation between the central black hole mass and properties of the host galactic spheroid observed at low redshift indicates that galaxy formation and massive black hole growth are fundamentally related \cite{ff05,fm00,gebhardt00,tremaine02}. A popular class of models to explain this correlation invokes black hole feedback in the form of winds or thermal energy input to regulate gas fueling \cite{sr98}. Such feedback effects might be even more important in the shallow potential wells of high-redshift galaxies \cite{li07}. An open problem is to understand the early growth rate of black holes, before they reach $\sim 10^6~M_{\odot}$, as a function of their cosmic environment, a process that ultimately leads to the formation of quasar-sized massive black holes at $z>6$ \cite{li07}. Due to the complexity involved, one needs to carry out high-resolution numerical simulations to investigate both the build-up of the first stellar systems and the early growth of massive black holes.

The internal structure of the first galaxies involves a wide range of physical scales, from the virial radius of the host dark matter halo to the much smaller scales such as the Jeans length of the cooling gas and the Bondi radius of the first massive black hole. Furthermore, the dynamical evolution of the first galaxies will involve turbulence and shocks that must be resolved. Some of the key questions are: What is the angular momentum content of the cold gas, and what processes affect angular momentum transport in the first rotationally-supported gas flow? Do global gravitational instabilities, such as bars-within-bars, give rise to rapid transport, and what are the consequences of rapid transport for the growth of the central object? What is the role of local gravitational instability and fragmentation? What is the impact of the radiation produced by the accreting black hole on the thermodynamic and chemical evolution of the first galaxy? At which point will a multi-phase interstellar medium form?

To simulate the build-up of the first stellar systems, we have to address the feedback from the very first stars on the surrounding intergalactic medium (IGM), and the formation of the second generation of stars out of material that was influenced by this feedback. There are a number of reasons why addressing the feedback from the first stars and understanding second-generation star formation is crucial:\\ {\it (i)} The first steps in the hierarchical build-up of structure provide us with a simplified laboratory for studying galaxy formation, which is one of the main outstanding problems in cosmology, given that we have successfully determined the parameters of the expanding background universe. The overall strategy is to start with cosmological initial conditions, follow the evolution up to the formation of a small number ($N<10$) of Pop~III stars, and trace the ensuing expansion of the supernova (SN) blast waves after they died together with the dispersal and mixing of the first heavy elements, towards the formation of second-generation stars out of enriched material \cite{greif07,wa07}.\\ {\it (ii)} The initial burst of Pop~III star formation may have been rather brief due to the strong negative feedback effects that likely acted to self-limit this formation mode \cite{gb06,ybh04}. Second-generation star formation, therefore, might well have been cosmologically dominant compared to Pop~III stars. Despite their importance for cosmic evolution, e.g., by possibly constituting the majority of sources for the initial stages of reionization at $z>10$, we currently do not know the properties, and most importantly the typical mass scale, of the second-generation stars that formed in the wake of the very first stars.\\ {\it (iii)} A subset of second-generation stars, those with masses below $\simeq 1~M_{\odot}$, would have survived to the present day. Surveys of extremely metal-poor Galactic halo stars therefore provide an indirect window into the Pop~III era by scrutinizing their chemical abundance patterns, which reflect the enrichment from a single, or at most a small multiple of, Pop~III SNe \cite{bc05,fjb07}. Stellar archaeology thus provides unique empirical constraints for numerical simulations, from which one can derive theoretical abundance patterns to be compared with the data.

Recently, observations of the highest-redshift quasars and galaxies to date have opened up a tantalizing window into the state of the universe at $z>7$ \cite{barton04,bi06,fan03,Iye06,kneib04,ricotti04,santos04,stanway04,yw04}. Studying the nature of star formation at even higher redshifts is of great interest, because the systems that we can directly observe at $z>7$ will already show the signature of previous stars. An intriguing example is the recently discovered $J-$band dropout HUDF-JD2, found through deep HST/VLT/{\it Spitzer} imaging, and interpreted to be a massive ($\sim 6\times 10^{11}~M_{\odot}$) post-starburst galaxy at $z>6.5$ \cite{mobasher05}. Since the observed stellar population in HUDF-JD2 is old, star formation could have begun already at $z\sim 15$, and this galaxy could thus contain the signature of first- and second-generation stars.

Existing and planned observatories, such as HST, Keck, VLT, and the {\it James Webb Space Telescope (JWST)}, planned for launch around 2013, yield data on stars and quasars less than a billion years after the Big Bang. The ongoing {\it Swift} gamma-ray burst (GRB) mission provides us with a possible window into massive star formation at the highest redshifts \cite{bl02,bl06,lr00}. Measurements of the near-IR cosmic background radiation, both in terms of the spectral energy distribution and the angular fluctuations provide additional constraints on the overall energy production due to the first stars \cite{dak05,fk06,kashlinsky05,msf03,sbk02}. Understanding the formation of the first galaxies is thus of great interest to observational studies conducted both at high redshifts and in our local Galactic neighborhood.

\section{Assembly of the First Galaxies}
How massive were the first galaxies, and when did they emerge? Theory predicts that dark matter (DM) halos containing a mass of $\sim 10^8~M_{\odot}$ and collapsing at $z\sim 10-15$ were the hosts for the first bona fide galaxies. These dwarf systems are special in that their associated virial temperature exceeds the threshold, $\sim 10^4~\rm{K}$, for cooling due to atomic hydrogen \cite{oh02}. These so-called `atomic-cooling halos' did not rely on the presence of molecular hydrogen to enable cooling of the primordial gas. In addition, their potential wells were sufficiently deep to retain photoionization heated gas, as well as SN shocked gas, in contrast to the shallow potential wells of minihalos \cite{dijkstra04,greif07,mfm02,mfr01}. These are arguably minimum requirements to set up a self-regulated process of star formation that comprises more than one generation of stars, and is embedded in a multi-phase interstellar medium. In the following, we discuss some of the key processes that govern the assembly of these first dwarf galaxies.

To avoid confusion, we would like to comment on a change in terminology. It has become evident that Pop~III star formation might actually consist of two distinct modes: one where the primordial gas collapses into a DM minihalo (see below), and one where the metal-free gas becomes significantly ionized prior to the onset of gravitational runaway collapse \cite{jb06}. We had termed this latter mode of primordial star formation `Pop~II.5' \cite{gb06,jb06,mbh03}. To more clearly indicate that both modes pertain to {\it metal-free} star formation, we here follow the new classification scheme suggested by Chris McKee (see McKee in these proceedings). Within this scheme, the minihalo Pop~III mode is now termed Pop~III.1, whereas the second mode (formerly `Pop~II.5') is now called Pop~III.2. The hope is that McKee's terminology will gain wide acceptance.

\subsection{Role of HD Cooling}
While the very first Pop~III stars (so-called Pop III.1), with masses of the order of $100~M_{\odot}$, formed within DM minihalos in which primordial gas cools by H$_2$ molecules alone, the HD molecule can play an important role in the cooling of primordial gas in several situations, allowing the temperature to drop well below $200~\rm{K}$ \cite{abn02,bcl02}. In turn, this efficient cooling may lead to the formation of primordial stars with masses of the order of $10~M_{\odot}$ (so-called Pop III.2) \cite{jb06}. In general, the formation of HD, and the concomitant cooling that it provides, is found to occur efficiently in primordial gas which is strongly ionized, owing ultimately to the high abundance of electrons which serve as catalyst for molecule formation in the early universe \cite{sk87}.

Efficient cooling by HD can be triggered within the relic H~{\sc ii} regions that surround Pop~III.1 stars at the end of their brief lifetimes, owing to the high electron fraction that persists in the gas as it cools and recombines \cite{jgb07,no05,yoshida07}. The efficient formation of HD can also take place when the primordial gas is collisionally ionized, such as behind the shocks driven by the first SNe or in the virialization of massive DM halos \cite{gb06,jb06,machida05,sv06}. In Figure~1, we show the HD fraction in primordial gas in four distinct situations: within a minihalo in which the gas is never strongly ionized, behind a $100~\rm{km}~\rm{s}^{-1}$ shock wave driven by a SN, in the virialization of a $3\sigma$ DM halo at redshift $z=15$, and in the relic H~{\sc ii} region generated by a Pop~III.1 star at $z\sim 20$ \cite{jb06}. Also shown is the critical HD fraction necessary to allow the primordial gas to cool to the temperature floor set by the CMB at these redshifts. Except for the situation of the gas in the virtually un-ionized minihalo, the fraction of HD becomes large quickly enough to play an important role in the cooling of the gas, allowing the formation of Pop~III.2 stars.

Figure~2 schematically shows the characteristic masses of the various stellar populations that form in the early universe. In the wake of Pop~III.1 stars formed in DM minihalos, Pop~III.2 star formation ensues in regions which have been previously ionized, typically associated with relic H~{\sc ii} regions left over from massive Pop~III.1 stars collapsing to black holes, while even later, when the primordial gas is locally enriched with metals, Pop~II stars begin to form \cite{bl03a,gb06}. Recent simulations confirm this picture, as Pop~III.2 star formation ensues in relic H~{\sc ii} regions in well under a Hubble time, while the formation of Pop~II stars after the first SN explosions is delayed by more than a Hubble time \cite{greif07,yoh07,yoshida07}.

\begin{figure}
\includegraphics[width=.35\textheight]{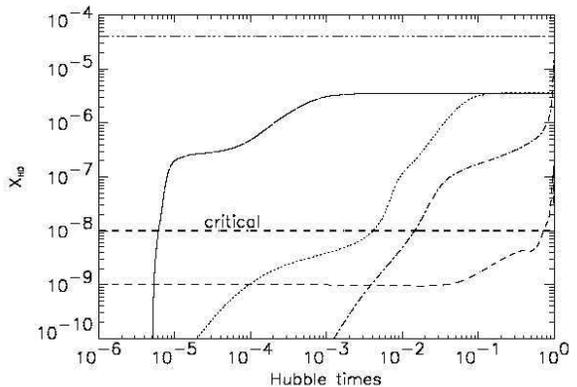}
\caption{Evolution of the HD abundance, $X_{\rm{HD}}$, in primordial gas which cools in four distinct situations. The solid line corresponds to gas with an initial density of $100~\rm{cm}^{-3}$, which is compressed and heated by a SN shock with velocity  $v_{\rm{sh}}=100~\rm{km}~\rm{s}^{-1}$ at $z=20$. The dotted line corresponds to gas at an initial density of $0.1~\rm{cm}^{-3}$ shocked during the formation of a $3\sigma$ halo at $z=15$. The dashed line corresponds to unshocked, un-ionized primordial gas with an initial density of $0.3~\rm{cm}^{-3}$ collapsing inside a minihalo at $z=20$. Finally, the dash-dotted line shows the HD fraction in primordial gas collapsing from an initial density of $0.3~\rm{cm}^{-3}$ inside a relic H~{\sc ii} region at $z=20$. The horizontal line at the top denotes the cosmic abundance of deuterium. Primordial gas with an HD abundance above the critical value, $X_{\rm{HD,crit}}$, denoted by the bold dashed line, can cool to the CMB temperature within a Hubble time.}
\end{figure}

\begin{figure}
\includegraphics[width=.35\textheight]{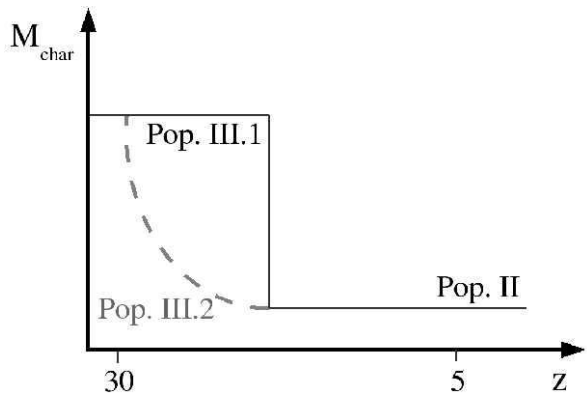}
\caption{Characteristic stellar mass as a function of redshift. Pop~III.1 stars, formed from unshocked, un-ionized primordial gas are characterized by masses of the order of $100~M_{\odot}$. Pop~II stars, formed in gas which is enriched with metals, emerged at lower redshifts and have characteristic masses of the order of $1~M_{\odot}$. Pop~III.2 stars, formed from ionized primordial gas, have characteristic masses reflecting the fact that they form from gas that has cooled to the temperature of the CMB. Thus, the characteristic mass of Pop~III.2 stars is a function of redshift, but is typically of the order of $10~M_{\odot}$.}
\end{figure}

\subsection{Radiative Feedback}
Due to their extreme mass scale, Pop~III.1 stars emit copious amounts of ionizing radiation, as well as a strong flux of H$_2$-dissociating Lyman-Werner (LW) radiation \cite{bkl01,schaerer02}. Thus, the radiation from the first stars dramatically influences their surroundings, heating and ionizing the gas within a few kiloparsec (physical) around the progenitor, and destroying the H$_2$ and HD molecules locally within somewhat larger regions \cite{abs06,awb07,ferrara98,jgb07,kitayama04,wan04}. Additionally, the LW radiation emitted by the first stars could propagate across cosmological distances, allowing the build-up of a pervasive LW background radiation field \cite{har00}.

The impact of radiation from the first stars on their local surroundings has important implications for the numbers and types of Pop~III stars that form. The photoheating of gas in the minihalos hosting Pop~III.1 stars drives strong outflows, lowering the density of the primordial gas and delaying subsequent star formation by up to $100~\rm{Myr}$ \cite{jgb07,wan04,yoshida07}. Furthermore, neighboring minihalos may be photoevaporated, delaying star formation in such systems as well \cite{as07,greif07,sir04,su06,whalen07}. The photodissociation of molecules by LW photons emitted from local star-forming regions will, in general, act to delay star formation by destroying the main coolants that allow the gas to collapse and form stars.

The photoionization of primordial gas, however, can ultimately lead to the production of copious amounts of molecules within the relic H~{\sc ii} regions surrounding the remnants of Pop~III.1 stars \cite{jb07,no05,oh02,rgs01}. Recent simulations tracking the formation of, and radiative feedback from, individual Pop~III.1 stars in the early stages of the assembly of the first galaxies have demonstrated that the accumulation of relic H~{\sc ii} regions has two important effects. First, the HD abundance that develops in relic H~{\sc ii} regions allows the primordial gas to re-collapse and cool to the temperature of the CMB, possibly leading to the formation of Pop~III.2 stars in these regions \cite{jgb07,yoh07}. Second, the molecule abundance in relic H~{\sc ii} regions, along with their increasing volume-filling fraction, leads to a large optical depth to LW photons over physical distances of the order of several kiloparsecs. The development of a high optical depth to LW photons over such short length-scales suggests that the optical depth to LW photons over cosmological scales may be very high, acting to suppress the build-up of a background LW radiation field, and mitigating negative feedback on star formation.

Figure~3 shows the chemical composition of primordial gas in relic H~{\sc ii} regions, in which the formation of H$_2$ molecules is catalyzed by the high residual electron fraction. Figure~4 shows the average optical depth to LW photons across the simulation box, which rises with time owing to the increasing number of relic H~{\sc ii} regions.

\begin{figure}
\includegraphics[width=.35\textheight]{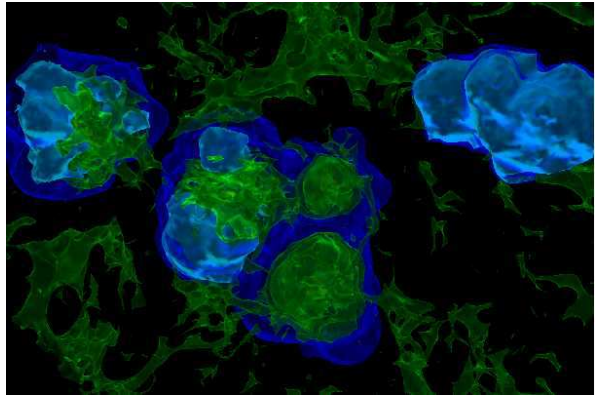}
\caption{The chemical interplay in relic H~{\sc ii} regions. While all molecules are destroyed in and around active H~{\sc ii} regions, the high residual electron fraction in relic H~{\sc ii} regions catalyzes the formation of an abundance of H$_2$ and HD molecules. The light and dark shades of blue denote regions with a free electron fraction of $5\times 10^{-3}$ and $5\times 10^{-4}$, respectively, while the shades of green denote regions with an H$_2$ fraction of $10^{-4}$, $10^{-5}$, and $3\times 10^{-6}$, in order of decreasing brightness. The regions with the highest molecule abundances lie within relic H~{\sc ii} regions, which thus play an important role in subsequent star formation, allowing molecules to become shielded from photodissociating radiation and altering the cooling properties of the primordial gas.}
\end{figure}

\begin{figure}
\includegraphics[width=.35\textheight]{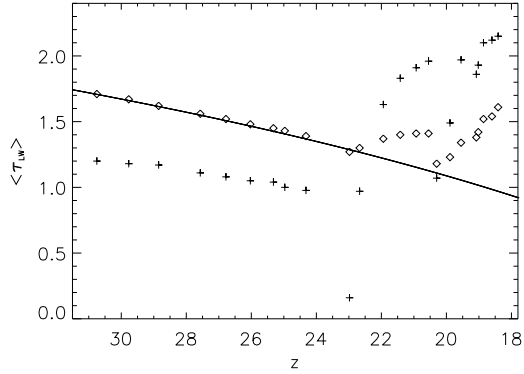}
\caption{Optical depth to LW photons averaged over two different scales, as a function of redshift. The diamonds denote the optical depth averaged over the entire cosmological box of comoving length $660~\rm{kpc}$, while the plus signs denote the optical depth averaged over a cube of $220~\rm{kpc}$ (comoving) per side, centered on the middle of the box. The solid line denotes the average optical depth that would be expected for a constant H$_2$ fraction of $2\times 10^{-6}$ (primordial gas), which changes only due to cosmic expansion.}
\end{figure}

\subsection{Massive Black Hole Growth}
The observations of quasars at redshift $z>6$ in the {\it Sloan Digital Sky Survey (SDSS)} suggest that some black holes in the early universe grew to have masses in excess of $10^9~M_{\odot}$ within the first billion years after the Big Bang \cite{fan04,fan06}. Many massive Pop~III stars may have collapsed directly to form black holes at the end of their brief lifetimes, thus providing the seed black holes which then accreted gas and grew to be the supermassive black holes powering the high-redshift quasars that are observed today \cite{heger03}.

In order for Pop~III remnant black holes to accrete $\sim 10^9~M_{\odot}$ of mass by $z\sim 6$, these objects would have to accrete gas at or near the Eddington rate for hundreds of millions of years. In turn, these black holes can only accrete at such high rates if the surrounding density exceeds $10^2~\rm{cm}^{-3}$ (see Figure~5) \cite{jb07}. However, as Figure~6 shows, the photoheating of the gas surrounding the first Pop~III stars in minihalos drives strong outflows from the central star, leaving a remnant black hole to reside in the middle of the evacuated minihalo with densities typically well below $1~\rm{cm}^{-3}$. Thus, the high accretion rates necessary to explain the presence of $10^9~M_{\odot}$ black holes at $z\sim 6$ are not possible for at least $50~\rm{Myr}$ after the formation of the black hole, at which time DM halo mergers and the re-collapse of the photoheated gas may raise the central density of the gas to above $10^2~\rm{cm}^{-3}$ \cite{jgb07}. Simulations tracking the later accretion history of such seed black holes have also found that it is difficult for these objects to grow fast enough to explain the {\it SDSS} quasar observations (see also Alvarez et al. in these proceedings) \cite{li07,pdc07}. Therefore, more exotic origins for these supermassive black holes may be required \cite{bl03b,bvr06,ln06}.

\begin{figure}
\includegraphics[width=.35\textheight]{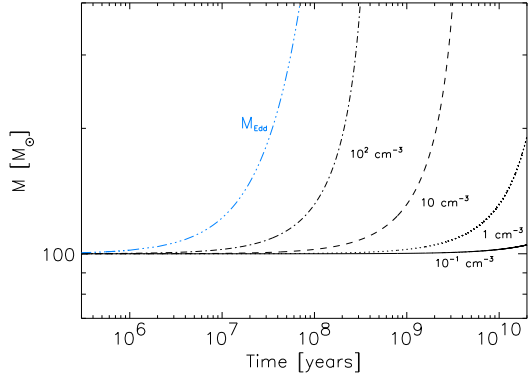}
\caption{The mass of an initially $100~M_{\odot}$ black hole as a function of time, assuming the black hole accretes gas at a temperature of $200~\rm{K}$ and at constant density. The solid, dotted, dashed, and dot-dashed lines show cases with densities of $0.1~\rm{cm}^{-3}$, $1~\rm{cm}^{-3}$, $10~\rm{cm}^{-3}$, and $100~\rm{cm}^{-3}$, respectively.  The triple dot-dashed line shows the mass of the black hole as a function of time, assuming that it accretes at the Eddington limit. Clearly, for the black hole to begin accreting at such a high rate while its mass is still of the order of $100~M_{\odot}$, the surrounding density must exceed $10^2~\rm{cm}^{-3}$.}
\end{figure}

\begin{figure}
\includegraphics[width=.35\textheight]{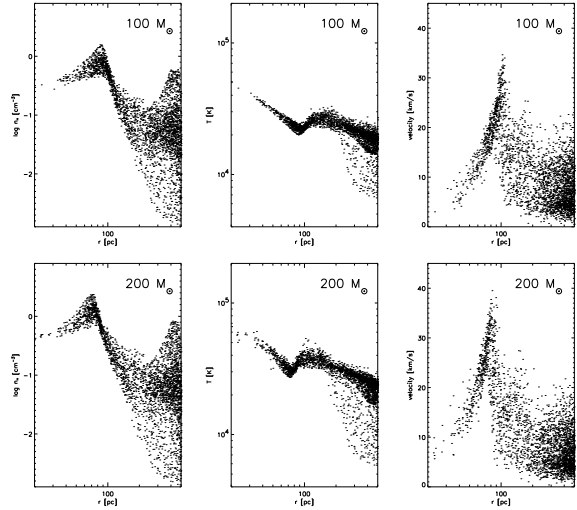}
\caption{The density, temperature and radial velocity of primordial gas heated and ionized by radiation from a $100~M_{\odot}$ and $200~M_{\odot}$ primordial star after their main-sequence lifetimes of $3~\rm{Myr}$ and $2~\rm{Myr}$, respectively. The central density drops to well below $1~\rm{cm}^{-3}$, which is insuffiecient to enable subsequent accretion at the Eddington rate (see Figure~5).}
\end{figure}

\subsection{The First Supernova Explosions}
Recent numerical simulations have indicated that primordial stars forming in DM minihalos typically attain $100~M_{\odot}$ by efficient accretion, and might even become as massive as $500~M_{\odot}$ \cite{bl04b,on07,op03,yoshida06}. After their main-sequence lifetimes of typically $2-3~\rm{Myr}$, stars with masses below $\simeq 100~M_{\odot}$ are thought to collapse directly to black holes without significant metal ejection, while in the range $140-260~M_{\odot}$ a pair-instability supernova (PISN) disrupts the entire progenitor, with explosion energies ranging from $10^{51}-10^{53}~\rm{ergs}$, and yields up to $0.5$ \cite{heger03,hw02}. Less massive primordial stars with a high degree of angular momentum might explode with similar energies as  hypernovae \cite{tun07,un02}.

The significant mechanical and chemical feedback effects exerted by such explosions have been investigated with a number of detailed calculations, but these were either performed in one dimension \cite{ky05,machida05,sfs04}, or did not start from realistic initial conditions \cite{byh03,nop04}. The most realistic simulation to date took cosmological initial conditions into account, and followed the evolution of the gas until the formation of the first minihalo at $z\simeq 20$ (see Figure~7) \cite{greif07}. After the gas approached the `loitering regime' at $n_{\rm{H}}\simeq 10^{4}~\rm{cm}^{-3}$, the formation of a primordial star was assumed, and a photoheating and ray-tracing routine determined the size and structure of the resulting H~{\sc ii} region (see Figure~8) \cite{jgb07}. An explosion energy of $10^{52}~\rm{ergs}$ was then injected as thermal energy into a small region around the progenitor, and the subsequent expansion of the SN remnant was followed until the blast wave effectively dissolved into the IGM. The cooling mechanisms responsible for radiating away the energy of the SN remnant, the temporal behavior of the shock, and its morphology could thus be investigated in great detail (see Figure~9).

\begin{figure}
\includegraphics[width=.35\textheight]{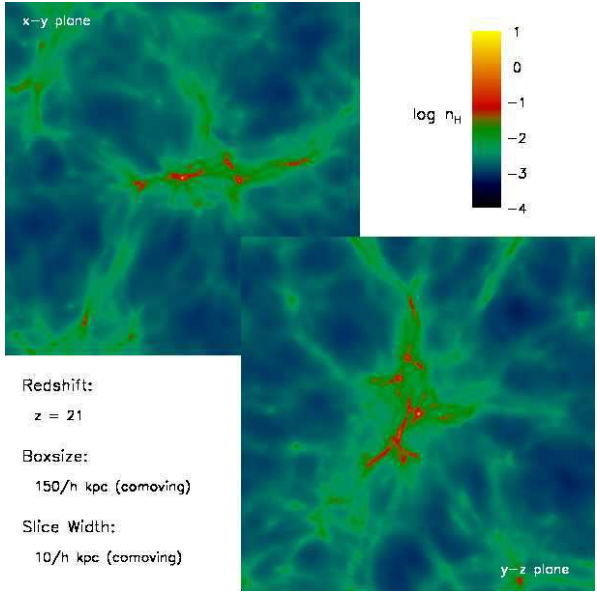}
\caption{Hydrogen number density averaged along the line of sight in a slice of $14~\rm{kpc}$ (comoving) around the first star, forming in a halo of total mass $M_{\rm{vir}}\simeq 5\times 10^{5}~M_{\odot}$ at $z\simeq 20$. Evidently, the host halo is part of a larger group of less massive minihalos, and subject to the typical bottom-up evolution of structure formation.}
\end{figure}

\begin{figure}
\includegraphics[width=.35\textheight]{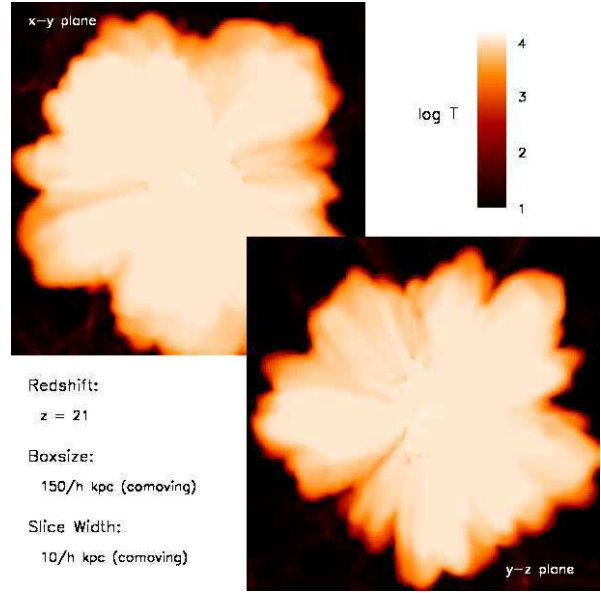}
\caption{Temperature averaged along the line of sight in a slice of $14~\rm{kpc}$ (comoving) around the star after its main sequence lifetime of $2~\rm{Myr}$. Ionizing radiation has penetrated nearby minihalos and extends up to $5~\rm{kpc}$ (physical) around the source, heating the IGM to roughly $2\times 10^{4}~\rm{K}$, while some high-density regions have effectively shielded themselves.}
\end{figure}

\subsubsection{Dynamical Evolution}
In analogy to present-day SNe, the evolution of SN remnants in the early universe can be decomposed into four physically distinct stages: free expansion (FE), Sedov-Taylor blast wave (ST), pressure-driven snowplow (PDS), and momentum-conserving snowplow (MCS) \cite{om88}. This allows the introduction of a simple analytic model, which reproduces the simulation results of \cite{greif07} relatively well (see Figure~9). It was found that the SN remnant propagates for a Hubble time at $z\simeq 20$ to a final mass-weighted mean shock radius of $2.5~\rm{kpc}$ (physical), roughly half the size of the H~{\sc ii} region, and sweeps up a total (gas) mass of $2.5\times 10^{5}~M_{\odot}$. The radial dispersion of the SN remnant increased dramatically once the shock left the host halo, although the bulk of the swept-up gas was expelled into the voids of the general IGM.

\begin{figure}
\includegraphics[width=.35\textheight]{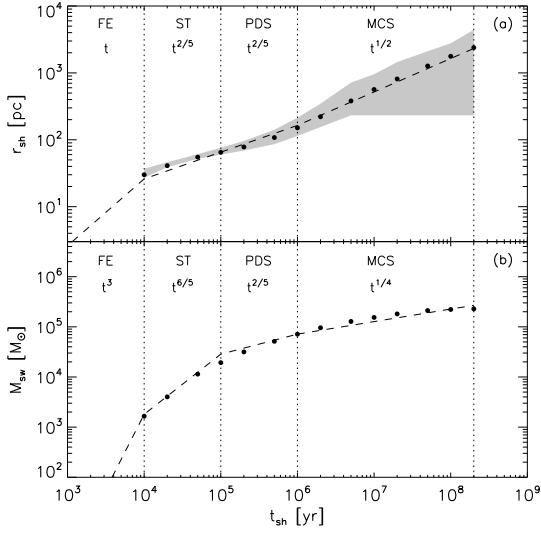}
\caption{The mass-weighted mean shock radius (Panel (a)), and swept-up (gas) mass (Panel (b)) of a $10^{52}~\rm{ergs}$ SN explosion in the high-redshift universe as a function of time (black dots), compared to the analytic model (dashed line). In the specific scenario considered, a final mass-weighted mean shock radius of $2.5~\rm{kpc}$ (physical) and a final swept-up mass of $2.5\times 10^{5}~M_{\odot}$ was found. The shaded region shows the radial dispersion of the shock, indicating that it increases dramatically once the SN remnant leaves the host halo and encounters the first neighbouring minihalos.}
\end{figure}

\subsubsection{Mechanical Feedback}
As confirmed in \cite{greif07}, highly energetic SN explosions are sufficient to entirely disrupt the host halo and evacuate their gaseous content \cite{mbh03}. This is directly related to their shallow potential wells, and was also found in previous calculations \cite{byh03,ky05}. Subsequent star formation does not ensue for at least a Hubble time at $z\simeq 20$, and in the specific scenario considered, the first Pop~II stars are expected to form out of enriched material at $z\simeq10$ \cite{greif07}.

Additional simulations in the absence of a SN explosion were performed to investigate the effect of photoheating and the impact of the SN shock on neighbouring minihalos (see Figure~10) \cite{greif07}. For the case discussed in this work, the SN remnant exerted positive mechanical feedback on neighbouring minihalos by shock-compressing their cores, while photoheating marginally delayed star formation. Although a viable theoretical possibility, secondary star formation in the dense shell via gravitational fragmentation was not observed, primarily due to the previous photoheating by the progenitor and the rapid adiabatic expansion of the post-shock gas \cite{machida05,mbh03,sfs04}.

\begin{figure}
\includegraphics[width=.35\textheight]{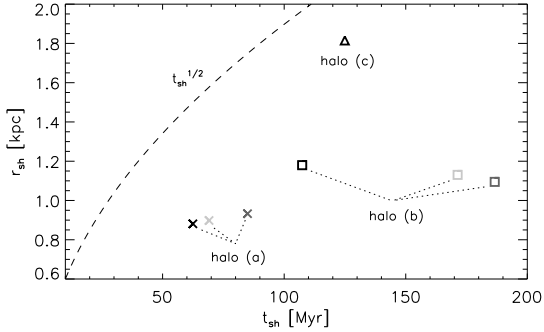}
\caption{The collapse times and distances from the SN progenitor for all star-forming minihalos affected by the SN shock. The shades of the symbols indicate their affiliation, i.e. black, dark grey and light grey symbols represent the no-feedback, photoheating-only and main simulation runs, respectively, while the shapes of the symbols denote the individual halos. For orientation, the dashed line shows the mass-weighted mean shock radius at late times according to Figure~9. In the cosmological realization considered, photoheating significantly delays star formation, while the SN shock compresses gas in neighbouring minihalos and slightly accelerates their collapse.}
\end{figure}

\subsubsection{Distribution of Metals}
The dispersal of metals by the first SN explosions transformed the IGM from a simple primordial gas to a highly complex medium in terms of chemistry and cooling, which ultimately enabled the formation of the first low-mass stars. However, this transition required at least a Hubble time, since the presence of metals became important only after the SN remnant had stalled and the enriched gas re-collapsed to high densities \cite{greif07}. Furthermore, the metal distribution was highly anisotropic, as the post-shock gas expanded into the voids in the shape of an `hour-glass', with a maximum extent similar to the final mass-weighted mean shock radius (see Figure~11) \cite{greif07}.

To efficiently mix the metals with all components of the swept-up gas, a DM halo of at least $M_{\rm{vir}}\simeq 10^{8}~M_{\odot}$ had to be assembled \cite{greif07}, and with an initial yield of $0.1$, the average metallicity of such a system would accumulate to $Z\simeq 10^{-2.5}Z_{\odot}$, well above any critical metallicity \cite{bl03a,bromm01,schneider06}. Thus, if energetic SNe were a common fate for the first stars, they would have deposited metals on large scales before massive galaxies formed and outflows were suppressed. Hints to such ubiquitous metal enrichment have been found in the low column density Ly$\alpha$ forest \cite{aguirre05,sc96,songaila01}, and in dwarf spheroidal satellites of the Milky Way \cite{helmi06}.

\begin{figure}
\includegraphics[width=.35\textheight]{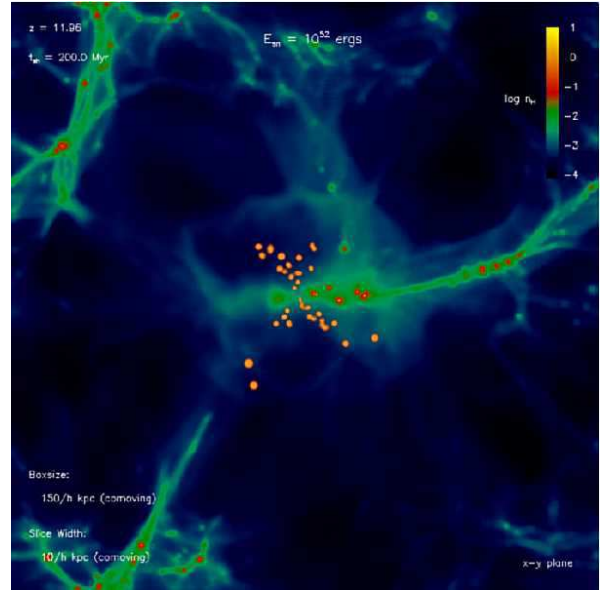}
\caption{Hydrogen number density averaged along the line of sight in a slice of $14~\rm{kpc}$ (comoving), overlayed with the distribution of all metal particles (bright orange) after $200~\rm{Myr}$, when the shock stalls. Most metals are dispersed into the voids of the general IGM, since the metal-rich interior expands adiabatically into the cavities created by the shock.}
\end{figure}

\section{Observational Constraints}
An increasing number of direct and indirect observations have become feasible during the last few years, due to rapid progress in observational methods and techniques. The most prominent among these concern the CMB optical depth to Thomson scattering, the near-IR background, high-redshift GRBs, and the possibility of scrutinizing the nature of the first stars by metals found in the oldest Galactic halo stars, dubbed `stellar archeology'. We here briefly discuss these promising observational avenues, and elaborate on their implications for the first stars and galaxies.

\subsubsection{Optical Depth to Thomson Scattering}
One of the most important constraints on early star formation is the CMB optical depth to Thomson scattering, recently revised to $\tau\simeq 0.09\pm 0.03$ after the {\it Wilkinson Microwave Anisotropy Probe (WMAP)} three-year data release \cite{spergel07}. Combined with the absence of the Gunn-Peterson trough in the spectra of high-redshift quasars, this measurement provides an integral constraint on the total ionizing photon production at $z>6$ \cite{alvarez06,wl03}.

In recent work, the contribution of the first stars to the ionizing photon budget was determined by semianalytic star formation rates based on the Sheth-Tormen formalism \cite{gb06}. It was found that a top-heavy primordial population (consisting of Pop~III.1 stars) must be terminated fairly rapidly in order to not overproduce the observed optical depth (see Figure~12). Such efficient feedback is possible via photoionization heating, which delays and possibly changes the mode of star formation in neighbouring minihalos, depending on the state of the minihalo collapse \cite{as07,mbh06,su06,whalen07,yoshida07}.

\begin{figure}
\includegraphics[width=.35\textheight]{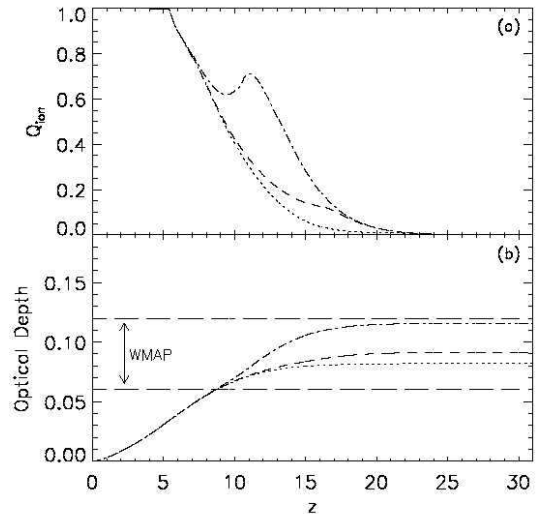}
\caption{Ionization histories (Panel (a)), and implied optical depths (Panel (b)) for a weak (dotted line), intermediate (dashed line), and strong (dot-dashed line) Pop~III.1 mode. For comparison, Panel~(b) includes the range of allowed values for the optical depth (long-dashed lines).}
\end{figure}

\subsubsection{Near-Infrared Background}
Another observation with the potential to yield valuable information on the nature of the first ionizing sources is the detection of an excess in the near-IR background of the order of $1~\rm{nW}~\rm{m}^{-2}~\rm{sr}^{-1}$ in the wavelength regime $1-2~\mu\rm{m}$ \cite{cambresy01}. If a fraction of this light is contributed by early stars, it corresponds to redshifted Ly$\alpha$ photons emitted between $z\simeq 7$ and $z\simeq 15$. Applying a semianalytic model of primordial star formation and deriving the total luminosity, it has been found that Pop~III.1 and Pop~III.2 combined must contribute at a level less than $10^{-3}~\rm{nW}~\rm{m}^{-2}~\rm{sr}^{-1}$ in order to avoid premature reionization \cite{gb06}. To determine what fraction is contributed by metal-free stars, one must analyze the fluctuation power of the IR background at scales small enough to capture their strong clustering \cite{dak05,kashlinsky05,msf03,sbk02}.

\subsubsection{High-redshift Gamma-ray Bursts}
GRBs are believed to originate in compact remnants (neutron stars or black holes) of massive stars, and their high luminosities make them detectable out to the edge of the visible universe \cite{cl00,lr00}. GRBs offer the opportunity to detect the most distant (and hence earliest) population of massive stars. In the hierarchical assembly process of DM halos, the first galaxies should have had lower masses (and lower stellar luminosities) than their low-redshift counterparts. Consequently, the characteristic luminosity of galaxies or quasars is expected to decline with increasing redshift. GRB afterglows, which already produce a peak flux comparable to that of quasars or starburst galaxies at $z\sim 1-2$, are therefore expected to outshine any competing source at the highest redshifts. As the electromagnetically-brightest explosions in the universe, GRBs should be detectable out to redshifts $z>10$ \cite{cl00,lr00}. High-redshift GRBs can be identified through IR photometry, based on the Ly$\alpha$ break induced by absorption of their spectrum at wavelengths below $1.216~\mu{\rm m}~[(1+z)/10]$. Follow-up spectroscopy of high-redshift candidates can then be performed on a 10-meter-class telescope. Recently, the ongoing {\it Swift} mission has detected a GRB originating at $z\simeq 6.3$ \cite{gehrels04,haislip06}, thus demonstrating the viability of GRBs as probes of the early universe.

Although the nature of the central engine that powers the relativistic jets of GRBs is still unknown, recent evidence indicates that long-duration GRBs trace the formation of massive stars \cite{bn00,bkd02,natarajan05,totani97,wijers98}, and in particular that long-duration GRBs are associated with Type Ib/c SNe \cite{stanek03}. Since the first stars in the universe are predicted to be predominantly massive \cite{abn02,bcl02,bl04a}, their death might give rise to large numbers of GRBs at high redshifts. In contrast to quasars of comparable brightness, GRB afterglows are short-lived and release $\sim 10$ orders of magnitude less energy into the surrounding IGM. Beyond the scale of their host galaxy, they have a negligible effect on their cosmological environment. However, the feedback from a single GRB or SN on the gas confined within the first galaxies could be dramatic, since the binding energy of most galaxies at $z>10$ is lower than $10^{51}~{\rm ergs}$ \cite{bl01}. Consequently, they are ideal probes of the IGM during the reionization epoch. Their rest-frame UV spectra can be used to probe the ionization state of the IGM through the spectral shape of the Gunn-Peterson (Ly$\alpha$) absorption trough, or its metal enrichment history through the intersection of enriched bubbles of SN ejecta from early galaxies \cite{fl03}.

What is the signature of GRBs that originates in metal-free, Pop~III progenitors? Simply knowing that a given GRB came from a high redshift is not sufficient to reach a definite conclusion as to the nature of the progenitor. For example, the currently highest-redshift GRB at $z\simeq 6.3$ clearly did not originate from a Pop~III progenitor, given the inferred level of metal enrichment in the host system of a few percent solar \cite{campana07}. Pregalactic metal enrichment was likely quite inhomogeneous, and we expect normal Pop~I and II stars to exist in galaxies that were already metal-enriched at these high redshifts \cite{bl06}. Pop~III and Pop~I/II star formation is thus predicted to have occurred concurrently at $z>5$. How is the predicted high mass-scale for Pop~III stars reflected in the observational signature of the resulting GRBs? Preliminary results indicate that circumburst densities are systematically higher in Pop~III environments. GRB afterglows will then be much brighter than for conventional GRBs. In addition, due to the systematically increased progenitor masses, the Pop~III distribution may be biased towards long-duration events. Figure~13 leads to the robust expectation that $\sim 10\%$ of all {\it Swift} bursts should originate at $z>5$. This prediction is based on the contribution from Pop~I/II stars which are known to exist even at these high redshifts. Additional GRBs could be triggered by Pop~III stars, with a highly uncertain efficiency.

\begin{figure}
\includegraphics[width=.35\textheight]{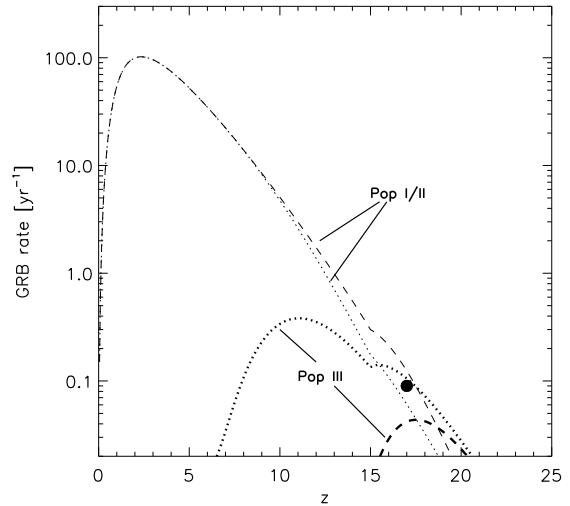}
\caption{Predicted GRB rate to be observed by {\it Swift} \cite{bl06}. Shown is the observed number of bursts per year, $dN_{\rm{GRB}}^{\rm{obs}}/d\ln (1+z)$, as a function of redshift. All rates are calculated with a constant GRB efficiency, $\eta_{\rm{GRB}}\simeq 2\times 10^{-9}~\rm{bursts}~M_{\odot}^{-1}$. {\it Dotted lines:} Contribution to the observed GRB rate from Pop~I/II and Pop~III for the case of weak chemical feedback. {\it Dashed lines:} Contribution to the GRB rate from Pop~I/II and Pop~III for the case of strong chemical feedback. {\it Filled circle:} GRB rate from Pop~III stars if these were responsible for reionizing the universe at $z\sim 17$.}
\end{figure}

\subsubsection{Stellar Archeology}
The discovery of extremely metal-poor stars in the Galactic halo has made studies of the chemical composition of low-mass Pop~II stars powerful probes of the conditions in which the first low-mass stars formed. While it is widely accepted that metals are required for the formation of low-mass stars, two general classes of competing models for the Pop~III -- Pop~II transition are discussed in the literature: {\it (i)} atomic fine-structure line cooling \cite{bl03a,ss06}; and {\it (ii)} dust-induced fragmentation \cite{schneider06}. Within the fine-structure model, C~{\sc ii} and O~{\sc i} have been suggested as main coolants \cite{bl03a}, such that low-mass star formation can occur in gas that is enriched beyond critical abundances of $\mbox{[C/H]}_{\rm{crit}}\simeq -3.5\pm 0.1$ and $\mbox{[O/H]}_{\rm{crit}}\simeq -3\pm 0.2$. The dust-cooling model, on the other hand, predicts critical abundances that are typically smaller by a factor of $10-100$.

Based on the theory of atomic line cooling \cite{bl03a}, a new function, the `transition discriminant' has been introduced:
 \begin{equation}
 D_{\rm{trans}}\equiv \log_{10}\left(10^{\mbox{[C/H]}}+0.3\times 10^{\mbox{[O/H]}}\right)\mbox{\ ,}
 \end{equation}
such that low-mass star formation requires $D_{\rm{trans}}>D_{\rm{trans,crit}}\simeq -3.5\pm 0.2$ \cite{fjb07}. Figure~14 shows values of $D_{\rm{trans}}$ for a large number of the most metal-poor stars available in the literature. While theories based on dust cooling can be pushed to accommodate the lack of stars with $D_{\rm{trans}}<D_{\rm{trans,crit}}$, it appears that the atomic-cooling theory for the Pop III -- Pop II transition naturally explains the existing data on metal-poor stars. Future surveys of Galactic halo stars will allow to further populate plots such as Figure~14, and will provide valuable insight into the conditions of the early universe in which the first low-mass stars formed.

\begin{figure}
\includegraphics[width=.35\textheight]{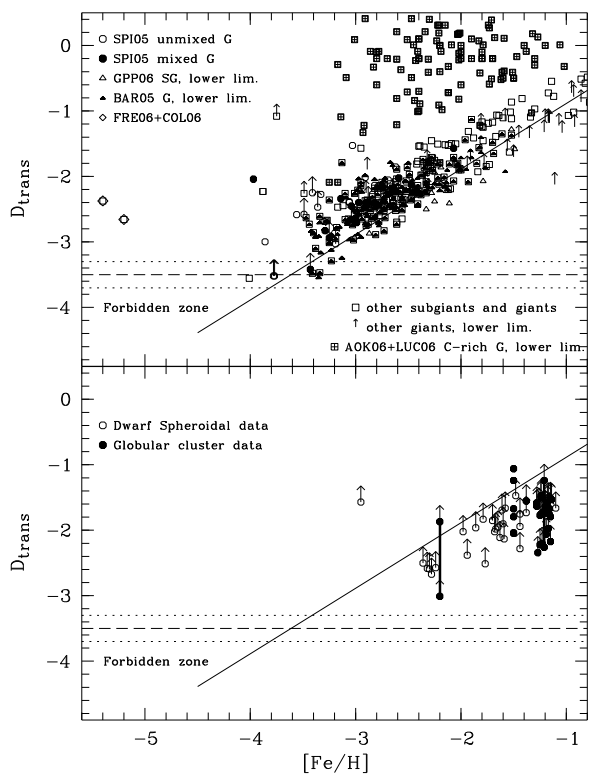}
\caption{Transition discriminant, $D_{\rm{trans}}$, for metal-poor stars collected from the literature as a function of [Fe/H]. {\it Top panel}: Galactic halo stars. {\it Bottom panel}: Stars in dSph galaxies and globular clusters. G indicates giants, SG subgiants. The critical limit is marked with a dashed line. The dotted lines refer to the uncertainty. The detailed references for the various data sets can be found in \cite{fjb07}.}
\end{figure}

\section{Outlook}
Understanding the formation of the first galaxies marks the frontier of high-redshift structure formation. It is crucial to predict their properties in order to develop the optimal search and survey strategies for the {\it JWST}. Whereas {\it ab-initio} simulations of the very first stars can be carried out from first principles, and with virtually no free parameters, one faces a much more daunting challenge with the first galaxies. Now, the previous history of star formation has to be considered, leading to enhanced complexity in the assembly of the first galaxies. One by one, all the complex astrophysical processes that play a role in more recent galaxy formation appear back on the scene. Among them are external radiation fields, comprising UV and X-ray photons, and possibly cosmic rays produced in the wake of the first SNe \cite{sb07}. There will be metal-enriched pockets of gas which could be pervaded by dynamically non-negligible magnetic fields, together with turbulent velocity fields built up during the virialization process. However, the goal of making useful predictions for the first galaxies is now clearly drawing within reach, and the pace of progress is likely to be rapid.

\begin{theacknowledgments}
V. B. acknowledges support from NSF grant AST-0708795 and NASA {\it Swift} grant NNX07AJ636. The simulations presented here were carried out at the Texas Advanced Computing Center (TACC). We are grateful to Paul Navr\'{a}til and Karla Vega at TACC for help with visualizations. T. H. G. would like to thank Paul Clark for suggestions which have improved the layout and content of this work.
\end{theacknowledgments}

\bibliographystyle{apj}

\end{document}